\RequirePackage{fix-cm}
\documentclass[twocolumn]{svjour3}          
\smartqed  
\usepackage{graphicx}
\usepackage{subfigure}
\usepackage{amssymb}
\usepackage{amsmath}
\usepackage{hyperref}
\usepackage{cite}
\usepackage{multirow}
 \journalname{Shock Waves}
\begin{document}

\title{Viscous solution of the triple-shock reflection problem
}


\author{S. She-Ming Lau-Chapdelaine         \and
        Matei I. Radulescu 
}

\authorrunning{S. SM. Lau-Chapdelaine, M. I. Radulescu} 

\institute{S.SM. Lau-Chapdelaine \at
            \email{slauc076@uottawa.ca}           
           \and
           M.I. Radulescu \at
			\email{matei@uottawa.ca}
			\\ \\University of Ottawa, \\Department of Mechanical Engineering,\\ 161 Louis Pasteur, Ottawa, Canada, K1N 6N5 \\
			\\
			The final publication is available at Springer via \url{http://dx.doi.org/10.1007/s00193-016-0674-8} including electronic supplementary material.
}

\date{Received: 19 May 2016 / Revised: 6 June 2016 / Accepted: 14 June 2016 / Published: September 2016}

\maketitle

\begin{abstract}
The reflection of a triple-shock configuration was studied numerically in two dimensions using the Navier-Stokes equations. The flow field was initialized using three shock theory, and the reflection of the triple point on a plane of symmetry was studied. The conditions simulated a stoichiometric methane-oxygen detonation cell at low pressure on time scales preceding ignition, when the gas was assumed to be inert.
Viscosity was found to play an important role on some shock reflection mechanisms believed to accelerate reaction rates in detonations when time scales are small. A small wall jet was present in the double Mach reflection and increased in size with Reynolds number, eventually forming a small vortex. Kelvin-Helmholtz instabilities were absent and there was no Mach stem bifurcation at Reynolds numbers corresponding to when the Mach stem had travelled distances on the scale of the induction length. Kelvin-Helmholtz instabilities are found to not likely be a source of rapid reactions in detonations at time scales commensurate with the ignition delay behind the Mach stem.

\keywords{Viscosity \and Shock wave \and Reflection}
\end{abstract}

\section{Introduction}
\label{intro}
\begin{figure}
	\centering
	\subfigure[\label{fig:regular reflection} Regular reflection]{
		\includegraphics[width=0.22\textwidth]{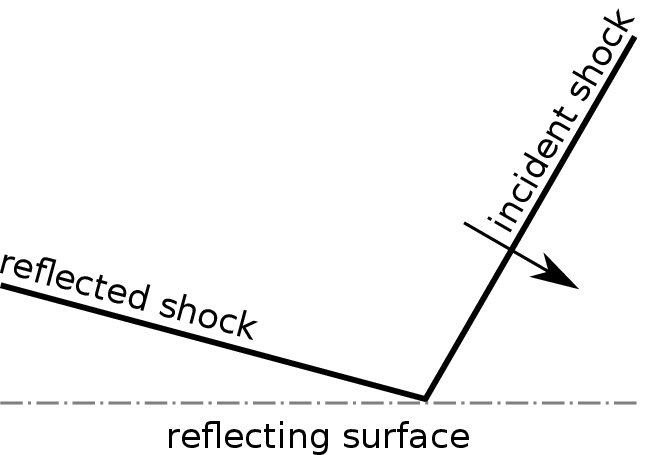}
	}
	\subfigure[\label{fig:irregular reflection} Mach reflection]{
		\includegraphics[width=0.22\textwidth]{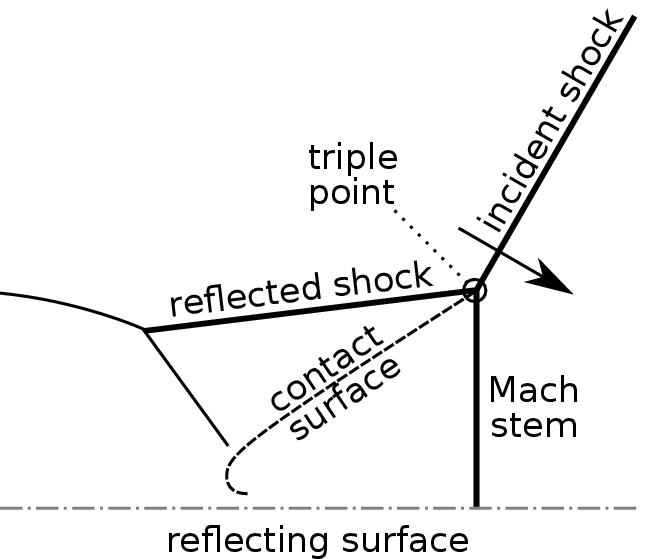}
	}
	\caption{\label{fig:shock_reflection} An incident shock that travels towards an oblique surface (direction of arrow) such as a wall or plane of symmetry, reflects and generally adopts one of two types of configuration: a regular reflection or a Mach (irregular) reflection}
\end{figure}

The problem of shock reflections is of importance to unsteady gas dynamics, including the study of detonation waves. When a shock wave reflects from a solid surface or a plane of symmetry, it usually adopts one of two forms: a regular reflection or a Mach reflection, illustrated in Fig. \ref{fig:shock_reflection}. 
Thorough reviews of shock reflection have been written by Hornung \cite{hornung_regular_1986} and Ben-Dor \cite{ben-dor_shock_2007}.

The Mach reflection is generally comprised of three shocks: the Mach (Mach stem), incident, and reflected (transverse) shocks, illustrated in Fig. \ref{fig:irregular reflection}. The Mach stem and incident shock compose the leading shock front and are joined at a kink called the triple point. The reflected wave emanates from the triple point and travels transversely behind the incident shock. A contact surface (sometimes called a slip line, or shear layer) separates gas shocked by the Mach stem from the gas shocked by the incident and reflected shock waves. The configuration of these discontinuities is determined by the incident shock (or Mach stem) strength, the angle between the incident and Mach shocks, and the isentropic exponent of the gas.

\begin{figure}
	\centering
	\includegraphics[width=0.5\textwidth]{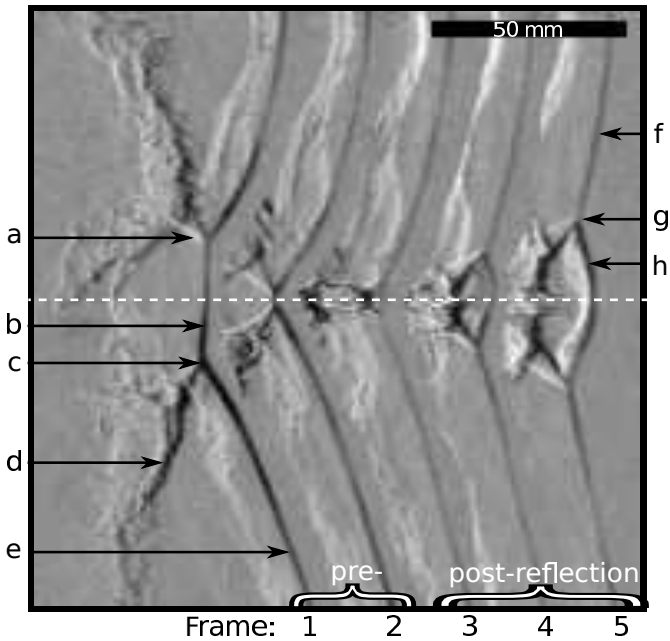}
	\caption{\label{fig:experiments} Superimposed schlieren photographs  (cropped, with grain extraction and stretched contrast) of two triple points colliding in a detonation, adapted from Bhattacharjee \cite{bhattacharjee_experimental_2013} (CH$_4$+2O$_2$, $\hat{p}_0=3.5$ kPa, $\hat{T}_0=300$ K, $\Delta \hat{t}= 11.53$ $\mu$s), dark features show a decrease in density and bright features show an increase from left to right, dotted line represents the axis of symmetry, a) reflected (transverse) shock, b) pre-reflection incident shock, c) triple point, d) contact surface, e) pre-reflection Mach stem, f) post-reflection incident shock, g) triple point, h) post-reflection Mach stem}
\end{figure}

Mach reflections are a natural part of the unstable (cellular) structure of detonations whose fronts have counter-travelling triple points, as seen experimentally in Fig. \ref{fig:experiments}; five successive photos have been pasted in place, meaning some details of each frame are covered by the preceding one. A detonation wave is shown travelling from left to right. The first two frames show a pair of triple points and their transverse shocks travelling towards each other, one from the bottom and the other from the top. The shocks then reflect off each other, forming a pair of triple points which move apart as seen in the next three frames. The path of the triple points over time form the cellular structure, whose dynamics have been well studied in the past \cite{subbotin_collision_1975,gamezo_formation_1999,austin_role_2003,strehlow_detonation_1971,radulescu_ignition_2005}.

Numerous studies have examined these shock reflections, seeking insight on the phenomena responsible for the creation of locally over-driven detonations, or re-initiation in cases of detonation failure. They have come upon a handful of candidates including: adiabatic shock compression, which heats the gas and exponentially reduces induction times \cite{teodorczyk_fast_1995,ohyagi_diffraction_2002,obara_reinitiation_2008,lau-chapdelaine_numerical_2013,lv_computational_2015}; jet formation, which may entrain combustion radicals from reacted zones in the rear to the zone behind the Mach stem \cite{sorin_detonation_2009,ziegler_simulations_2012,lau-chapdelaine_numerical_2013,bhattacharjee_experimental_2013,maley_influence_2015}, where subsequent mixing with unburnt gases may increase reaction rates; Richtmyer-Meshkov instabilities, which arise from the interaction between pressure waves and density gradients (from reactions), accelerating mixing \cite{bhattacharjee_experimental_2013}; and Kelvin-Helmholtz instabilities along shear layers, which may also accelerate mixing \cite{bhattacharjee_experimental_2013}.
While these studies have been able to capture and study reactive shock reflections and have clarified the importance of certain phenomena over others, much still remains to be clarified.

Experiments have suffered from the disparity of scales associated with the problem. The cellular structure of detonations can measure hundreds of induction lengths, which can be a hundred times longer than the energy deposition zone. The difficulty this causes in measuring the flow field is aggravated by the stochastic nature of detonations. Numerical simulations have suffered from different issues. For example, inviscid numerical simulations of shock reflections have predicted bifurcation of the Mach stem under certain conditions \cite{mach_mach_2011,glaz_detailed_1985}, caused by a strong jet and vortex, leading to an acceleration of the Mach stem and creation of new cells, however, this aspect of shock reflection has been absent in experiments. Simulations using the Euler equations do not converge and increasing resolution only causes a rise in non-physical phenomena \cite{sun_note_2003}. The validity of the inviscid flow assumption has come into question recently \cite{radulescu_hydrodynamic_2007} as mechanisms for turbulent mixing have been suggested to significantly increase reaction rates behind the detonation front.

This study will examine the importance of diffusion on the triple-shock reflection problem for conditions relevant to detonations. This will be done numerically by imposing the ideal three shock solution as the initial conditions of the domain, unique to this work. The triple-shock reflection process in hydrogen detonations have been described in detail \cite{sharpe_transverse_2001,hu_structure_2005}, but have not considered viscosity. In this case, the effects of viscosity will be resolved down to the shock thickness as done by Ziegler \cite{ziegler_simulations_2012}. While the role of turbulence will not be addressed directly, the simulations will offer insight on some turbulence-generating mechanisms such as Kelvin-Helmholtz instabilities and the wall jetting effect. The window of interest will cover the initial shock reflection at the tip of a detonation cell and its subsequent growth before ignition of the gas behind the Mach stem. This study extends previously presented work \cite{lau-chapdelaine_viscous_2015} with the inclusion of heat diffusion and an improved viscous model.

\section{Model}
\newcommand{\pde}[2]{\frac{\partial #1}{\partial #2}}
The flow is modelled using the compressible Navier-Stokes equations in two dimensions with momentum and heat diffusion
\begin{equation}
\begin{aligned}
\frac{\partial \hat{\rho}}{\partial \hat{t}} + \vec{\nabla} \cdot (\hat{\rho} \vec{\hat{u}}) = 0
\\
\frac{\partial	(\hat{\rho} \vec{\hat{u}})}{\partial \hat{t}} + \vec{\nabla} \cdot (\hat{\rho} \vec{\hat{u}} \vec{\hat{u}}) + \vec{\nabla} \hat{p} - \vec{\nabla} \cdot \vec{\vec{\hat{\tau}}} = 0
\\
\frac{\partial (\hat{\rho} \hat{E})}{\partial \hat{t}} + \vec{\nabla} \cdot \left((\hat{\rho} \hat{E} + \hat{p}) \vec{\hat{u}} - \vec{\hat{u}}\cdot \vec{\vec{\hat{\tau}}} + \vec{\hat{q}}\right) = 0
\end{aligned} \label{eq:Navier-Stokes}
\end{equation}
where
$\rho$ is the density, $p$ is the pressure, $u$ is the velocity, $t$ is time, $\vec{\vec{\tau}}$ is the shear stress tensor
\begin{equation}
\vec{\vec{\hat{\tau}}} = 2 \hat{\mu} \left( \frac{1}{2}\left(\pde{\hat{u}_{i}}{\hat{x}_{j}} + \pde{\hat{u}_{j}}{\hat{x}_{i}} \right) - \frac{1}{3} \delta_{ij} \vec{\nabla} \cdot \vec{\hat{u}} \right) \vec{e}_{i} \vec{e}_j, \notag
\end{equation}
in Einsteinian notation with viscosity $\mu$. The circumflex accent represents dimensional values. The bulk (second) viscosity is neglected. The total energy $E$ is
\begin{equation}
\hat{E} = \frac{\hat{p}}{\hat{\rho} (\gamma-1)} + \frac{1}{2} \vec{\hat{u}}\cdot \vec{\hat{u}}, \notag
\end{equation}
with constant isentropic exponent $\gamma$ for a calorically perfect gas, and $\vec{q}$ is the heat conduction
\begin{equation}
\vec{\hat{q}} = - \hat{k} \vec{\nabla} \hat{T} \notag
\end{equation}
with thermal conductivity $k$ and temperature $T$. The perfect gas equation of state is
\begin{equation}
\hat{p} = \hat{\rho} \hat{R} \hat{T} \notag
\end{equation}
with specific gas constant $R$. Viscosity is modelled using Sutherland's law with a coefficient of zero, and the Prandtl number is held constant so that
\begin{equation}
\hat{\mu} = \frac{\mu_{\mathrm{ref}}}{\sqrt{T_{\mathrm{ref}}}}\sqrt{\hat{T}} \mathrm{\ \ \ \ \ and\ \ \ \ \ } \hat{k} = \frac{k_{\mathrm{ref}}}{\sqrt{T_{\mathrm{ref}}}}\sqrt{\hat{T}}.  \label{eq:viscosity}
\end{equation}

The results will be compared to simulations of the Euler equations, which neglect the diffusive terms $\vec{\vec{\tau}}$ and $\vec{q}$, but have the same maximal resolution. The solutions to the Euler equations are ``inviscid", but are still subject to artificial diffusion from discretization of the equations.

The system of equations is non-dimensionalised by
\begin{equation}
\begin{aligned}
\rho = \frac{\hat{\rho}}{\hat{\rho}_{\mathrm{ref}}}, \mathrm{\ }
p = \frac{\hat{p}}{\hat{p}_{\mathrm{ref}}}, \mathrm{\ } 
T = \frac{\hat{T}}{\hat{T}_{\mathrm{ref}}}, \mathrm{\ } 
x_i = \frac{\hat{x}_i}{\hat{\lambda}_{\mathrm{ref}}}, \mathrm{\ }
u_i = \frac{\hat{u}_i}{\sqrt{\frac{\hat{\rho}_{\mathrm{ref}}}{\hat{p}_{\mathrm{ref}}}}}, 
\\
t = \frac{\hat{t}}{\hat{\lambda}_{\mathrm{ref}} \sqrt{\frac{\hat{p}_{\mathrm{ref}}}{\hat{\rho}_{\mathrm{ref}}}}}, \mathrm{\ }
\mu = \frac{\hat{\mu}}{\sqrt{\hat{\rho}_{\mathrm{ref}}\hat{p}_{\mathrm{ref}}} \hat{\lambda}_{\mathrm{ref}}}, \mathrm{\ }
k = \frac{\hat{k}}{\hat{\rho}_{\mathrm{ref}} \hat{p}_{\mathrm{ref}}^{\frac{3}{2}} \hat{T}_{\mathrm{ref}}^{-1} \hat{\lambda}_{\mathrm{ref}}}. 
\end{aligned} \notag
\end{equation}
The reference state (subscript ref) is chosen to be the unshocked state with mean free path \cite{vincenti_introduction_1965}
\begin{equation}
\hat{\lambda}_{\mathrm{ref}} = \frac{\hat{R} \hat{T}_{\mathrm{ref}}}{\sqrt{2}\pi \hat{d}^2 \hat{N}_{\mathrm{A}} \hat{p}_{\mathrm{ref}}} \notag
\end{equation}
where $\hat{N}_{\mathrm{A}}$ is Avogadro's constant and $\hat{d}$ is the molecular size. The viscosity model \eqref{eq:viscosity} at this reference state, along with the omission of the bulk viscosity, which can be quite significant in some gasses \cite{thompson_compressible-fluid_1988,cramer_numerical_2012}, under-predicts the viscosity in high temperature regions (\textit{e.g.}, behind the Mach stem). This will likely have little qualitative impact on the results, even strengthening arguments made in the discussion.

The gas is considered chemically frozen since the shock reflection is being examined on time scales less than the induction time, prior to noticeable thermicity.

\section{Numerical method}
\label{sec:1}
The compressible Navier-Stokes equations \eqref{eq:Navier-Stokes} in two dimensions were solved using \verb|mg|, a computational package developed by Falle \cite{falle_self-similar_1991,falle_upwind_1996}. This was done with a second-order accurate exact Godunov scheme for the convective terms, and diffusive terms were solved explicitly.

\begin{figure}
\centering
    \includegraphics[scale=1]{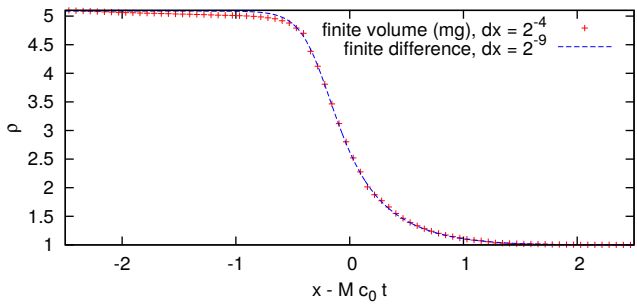}
    \caption{One-dimensional Navier-Stokes shock profile of density shows that the shock thickness is nearly converged at $t=1$ and a resolution of 16 grids per mean free path}
    \label{fig:shock_profile}
\end{figure}

A resolution study was performed on an initially ideal one-dimensional shock with Mach number $M=4.363$ through a quiescent gas with isentropic exponent $\gamma=1.358$, relating to the experimental strength of the pre-reflection incident shock of Fig. \ref{fig:experiments} travelling through the unshocked gas. Fig. \ref{fig:shock_profile} compares results from the computational package \verb|mg| at $t=1$ to the steady-state Navier-Stokes shock profile, integrated using first order finite differences (converged at the given resolution). The integration began with a relative deviation of $10^{-9}$ from the ideal post-shock conditions and shot towards the pre-shock state. The figure shows that the shock thickness is converged at a maximum resolution of approximately $16$ grids per mean free path and that $t=1$ is sufficient for the ideal shock to reach its steady viscous form.

A rectangular domain of size 240 by 180 mean free paths was modelled, totalling in width about one hundredth of the scale bar of Fig. \ref{fig:experiments}. The domain was covered by a Cartesian mesh with 60 by 45 grid points with six additional levels of adaptive mesh-refinement, each doubling the resolution when differences between flow properties exceeded a tolerance of 0.01 between levels. The bottom wall used a symmetry boundary condition while the remainder were set to have zero normal gradients; all were parallel to the mesh. The triple point was positioned at the origin, equidistant from the left and right boundaries and $4.5$ mean free paths above the reflecting surface (bottom wall) to let the viscous shock structure fully develop prior to reflection. This means the reflection occurs $t=1.047$ after initiation. The domain was sized in the $y$-direction to allow enough room for the reflection to grow to the desired size before boundary error reached the reflected triple point. Results at larger times were obtained using a larger domain at the same resolution, with the right boundary extending further from the origin than the left boundary.

Past numerical shock reflection studies have usually considered the interaction of a single shock front with a reflecting wedge, however, the reflection of a triple point from a symmetry boundary has been modelled instead. This better reproduces the geometry of cell collision and removes certain numerical difficulties associated with the creation of reflection boundaries in Cartesian grids \cite{mach_bifurcating_2011,lau-chapdelaine_non-uniqueness_2013} and reduces early-reflection artifacts, which can affect the reflection at later times \cite{lau-chapdelaine_non-uniqueness_2013,previtali_unsteady_2015}.

\begin{figure}
\centering
    \includegraphics[width=0.5\textwidth]{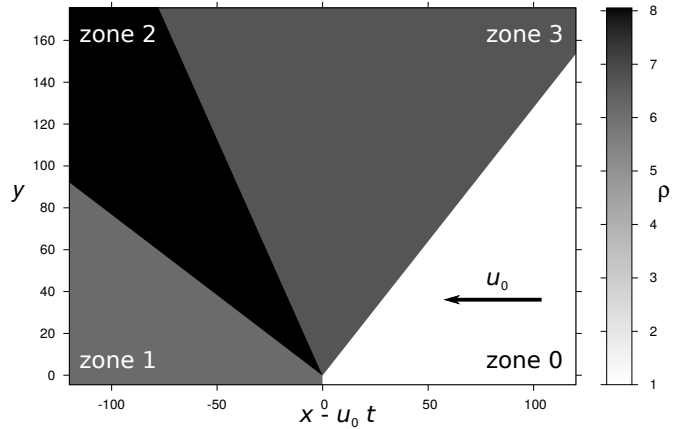}
    \caption{Density plot of initial conditions}
    \label{fig:t=0}
\end{figure}

Conditions approximating the second frame of Fig. \ref{fig:experiments} (top triple point) were initially imposed onto the domain. These correspond to the experimental results of a detonation passing through stoichiometric methane-oxygen performed by Bhattacharjee \cite{bhattacharjee_experimental_2013}. The initial conditions were calculated using the ideal three-shock solution \cite{hornung_regular_1986,ben-dor_shock_2007} with an incident shock of strength $M=4.363$ normal to the reflecting boundary and an angle of $141.95 ^{\circ}$ between the incident and Mach shocks. The triple point's frame of reference in the $x$-direction was used, with the unshocked gas travelling at a velocity $u_0=5.085$ to the left, reducing the required domain size. A constant isentropic exponent $\gamma=1.358$ was assumed, corresponding to the unshocked conditions of stoichiometric methane-oxygen at $300$ K and $3.5$ kPa. The length scale under these conditions $\hat{\lambda}_{\mathrm{ref}} \approx 2.02 \times 10 ^{-6} \mathrm{\ m}$ is calculated using $\hat{d} = 3.63 \times 10^{-10}$ m as the mole-averaged molecular size for a stoichiometric mixture of methane \cite{flynn_lennard-jones_1962} and oxygen \cite{hirschfelder_molecular_1954}. The initial conditions are tabulated in table \ref{tab:IC}, and shown in Fig. \ref{fig:t=0}. 

The reflected triple shock can also be predicted using three shock theory, by keeping the isentropic exponent and angle between the incident and Mach shocks constant, but imposing the pre-reflection Mach stem strength as the post-reflection incident shock strength. The ideal triple shocks and their reflections referred to in this text are calculated assuming a constant isentropic exponent. The transport properties behind the various shocks are then estimated using a chemical kinetics code, Cantera \cite{goodwin_cantera_2016} with the GRI-3.0 mechanism \cite{smith_gri_1999}, assuming no change in chemical composition.

\begin{table}[]
\centering
\caption{\label{tab:IC}Initial conditions, zones refer to Fig. \ref{fig:t=0}}
\begin{tabular}{c|c|c|c|c}
\multicolumn{1}{c}{} & Unshocked & Incident & Reflected & Mach \\ 
\multicolumn{1}{c}{} & (zone 0) & (zone 1) & (zone 2) & (zone 3) \\ 
\hline \rule{0pt}{2.2ex} 
$p$ & 1 & 21.7767 & 41.0169 & 41.0169 \\ 
$\rho$  & 1 & 5.08848 & 8.05256 & 5.69144 \\ 
$u$  & -5.08513 & -0.999342 & -1.71771 & -0.562407 \\ 
$v$ & 0 & 0 & -0.93582 & -3.5399 \\ 
\hline 
\end{tabular} 
\end{table}

\section{Results}
\label{sec:2}

\begin{figure*}[t]
\centering
    \includegraphics[scale=1]{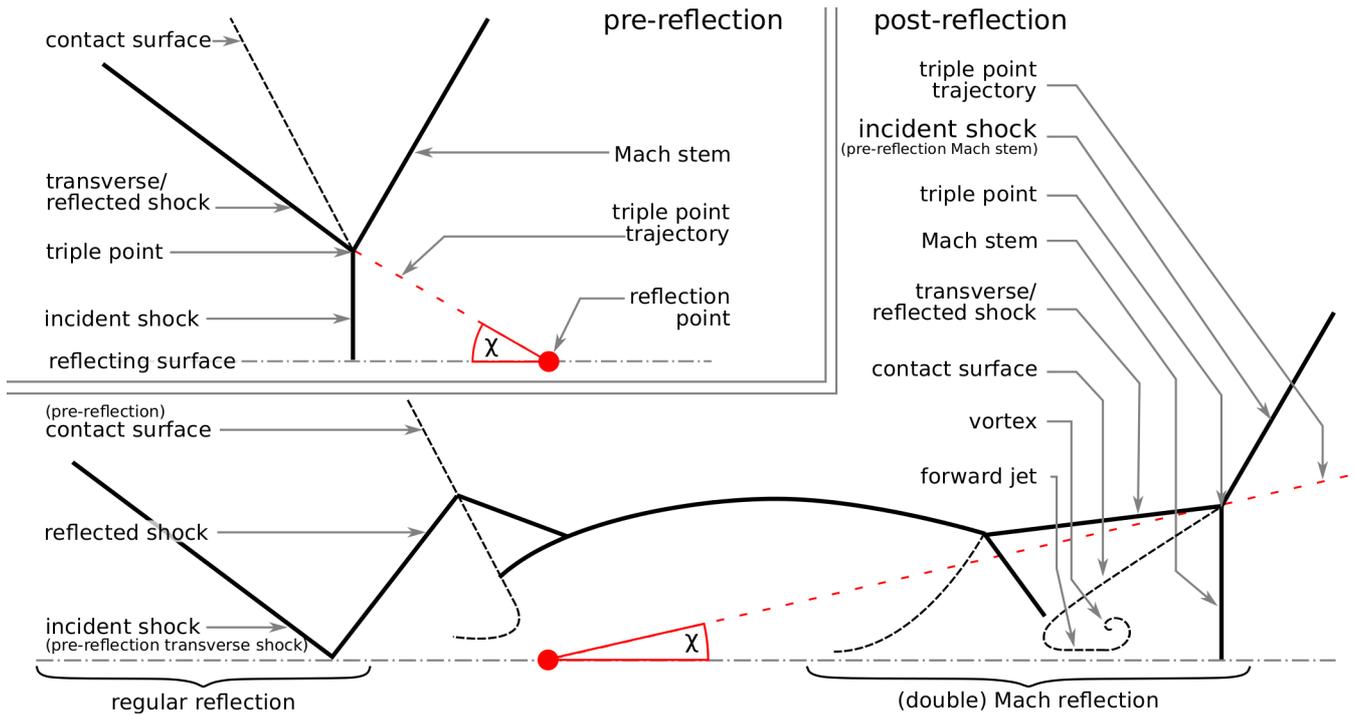}
    \caption{Illustration of the three shock structure before and after reflection}
    \label{fig:triple_shock_reflection}
\end{figure*}

The triple-shock reflection is illustrated and labelled in Fig. \ref{fig:triple_shock_reflection}. The triple point initially travels along the trajectory denoted by the red dashed line until it reaches the reflection point. The shock wave known as the pre-reflection Mach stem then continues as the post-reflection incident shock on the right side and forms a new Mach stem as it interacts with the reflecting surface as a double Mach reflection. The pre-reflection transverse shock also interacts with the surface and forms a regular reflection on the left side. The prefixes ``pre-reflection'' and ``post-reflection'' when used will distinguish whether the features being referred to are those before or after, respectively, the triple point has reached the reflection point as they otherwise may share the same name (\textit{e.g.}, the pre-reflection incident shock refers to the vertical feature in the ``pre-reflection'' box of figure \ref{fig:triple_shock_reflection}).

\begin{figure*}[]
\centering
\subfigure[\label{fig:r,t=17,viscous} Density, viscous, $\mathrm{Re} \approx 100$, $t=17$]{
        \includegraphics[width=\textwidth]{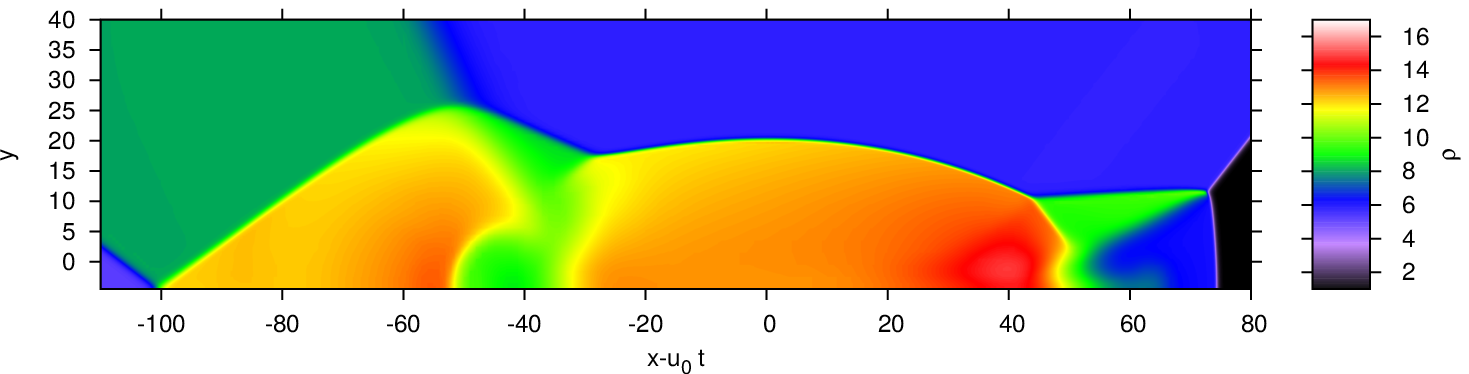}
} \\
\subfigure[\label{fig:r,t=17,invsicid} Density, inviscid,  $t=17$]{
        \includegraphics[width=\textwidth]{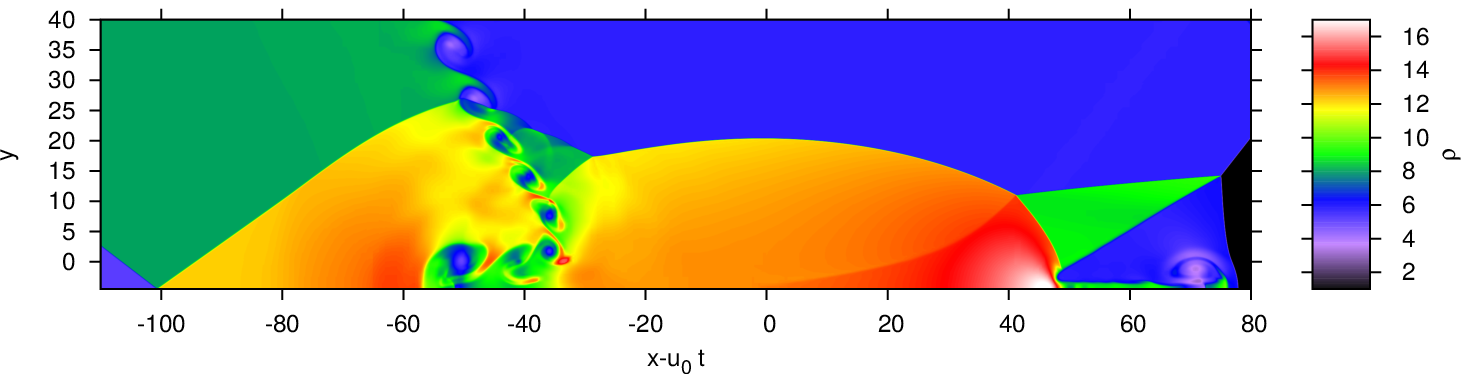}
} \\
\subfigure[\label{fig:T,t=17,viscous} Temperature, viscous, $\mathrm{Re} \approx 100$, $t=17$]{
        \includegraphics[width=0.47\textwidth]{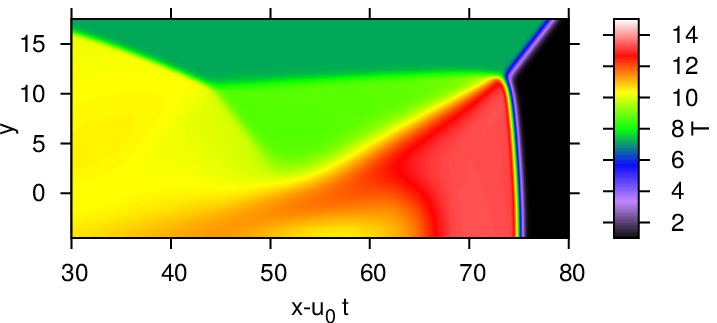}
}
\subfigure[\label{fig:T,t=17,inviscid} Temperature, inviscid,  $t=17$]{
        \includegraphics[width=0.47\textwidth]{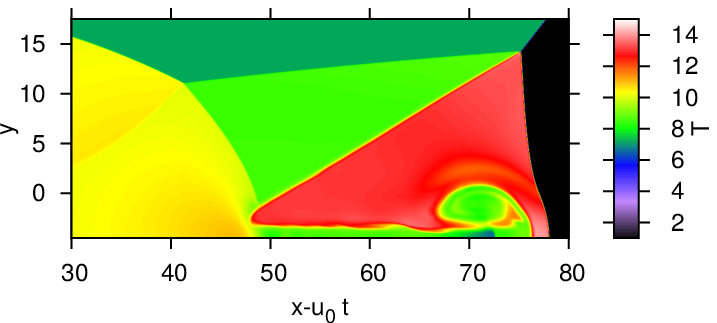}
} \\
\subfigure[\label{fig:T,t=120,viscous} Temperature, viscous, $\mathrm{Re} \approx 800$, $t=120$]{
        \includegraphics[width=0.47\textwidth]{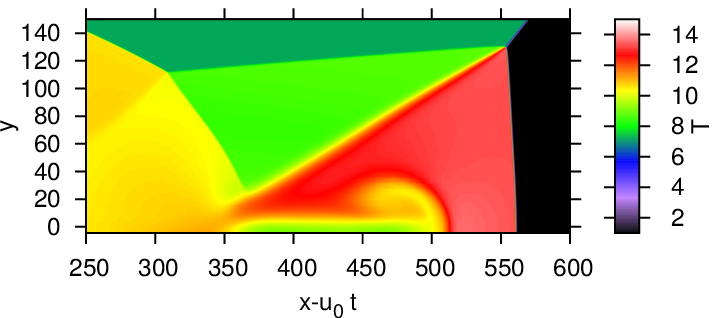}
}
\subfigure[\label{fig:T,t=120,inviscid} Temperature, inviscid, $t=120$]{
        \includegraphics[width=0.47\textwidth]{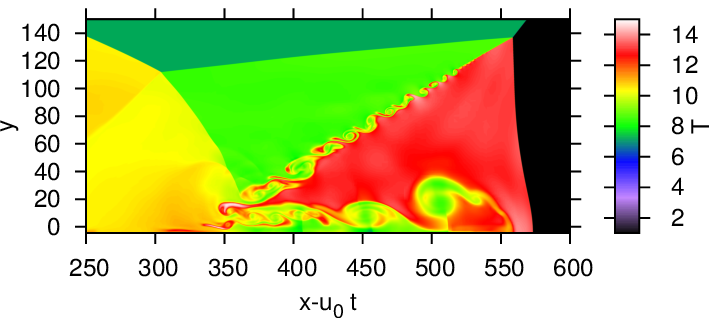}
}
	\caption{\label{fig:results} Simulations results cropped to areas of interest for density $\rho$ and temperature $T$; ranges kept consistent for comparison}
\end{figure*}

This is observed in the simulation results of Fig. \ref{fig:results} and as electronic supplementary material (Online Resources 1 and 2 for the viscous and inviscid videos, respectively, with frame rate $\Delta t = 1$). The density profile of the viscous reflection in Fig. \ref{fig:r,t=17,viscous} is shown at the approximate induction time of gas behind the post-reflection Mach stem ($t=17$), approximated assuming a constant volume combustion with the ideal post-reflection Mach state as initial conditions, computed using Cantera \cite{goodwin_cantera_2016} and the GRI 3.0 mechanism \cite{smith_gri_1999}. The double Mach reflection formed on the right is seen at $x - u_0 t \approx 75$,  and the regular reflection formed on the left is seen at $x - u_0 t \approx -100$. The key features of the Mach reflection will be reviewed first, followed by those of the regular reflection.

\subsection{Mach Reflection}

A double Mach reflection propagating through the unshocked gas arises from the reflection of the pre-reflection Mach stem (now considered the incident shock) with the bottom wall. The triple point follows a trajectory of $5.92^{\circ} \pm 0.10^{\circ}$ above the horizontal in the lab frame of reference (calculated every $\Delta t = 1$ in the range $8 \le t \le 17$), which is below the inviscid simulation's $6.32^{\circ} \pm 0.04^{\circ}$.

There is little evidence of Kelvin-Helmholtz instabilities along the contact surface behind the Mach stem in the inviscid case (Fig. \ref{fig:r,t=17,invsicid}), and none in the viscous case (Fig. \ref{fig:r,t=17,viscous}). The contact surface jets towards the Mach stem along the reflecting wall, but this wall jet does not reach the Mach stem in the viscous simulation. A close-up of the Mach stem's temperature profiles in Fig. \ref{fig:T,t=17,viscous} and \ref{fig:T,t=17,inviscid} shows that the removal of viscous terms has the wall jet curl into a vortex as the jet reaches the Mach stem, causing it to bulge, but remaining unbifurcated \cite{mach_mach_2011}.

The temperature profiles near the Mach reflection show that the hottest regions are found behind the Mach stem, especially at its foot. The vortex behind the Mach stem, present only in the inviscid simulation, is composed mostly of cool gas near the temperature of the gas shocked by the reflected wave. 

When the simulation time is extended to the ignition time behind the post-reflection transverse shock, shown in Fig. \ref{fig:T,t=120,viscous} at $t=120$, a comparison of the viscous solution with the earlier time shows that temperature in the centre of the wall jet decreased, while the temperature behind the Mach stem slightly increased, and a vortex began to form at the end of the wall jet. The shock that joins the contact surface and transverse wave (making this a double Mach reflection) has become sharper and its triple point, the kink along the transverse wave, has become much more distinguished. The inviscid solution (Fig. \ref{fig:T,t=120,viscous}) develops Kelvin-Helmholtz instabilities along the contact surface which grow through to the wall jet. Mach stem bulging is reduced, relative to the Mach stem height.

\subsection{Regular reflection}

The regular reflection seen on the left of Fig. \ref{fig:r,t=17,viscous} is caused by the reflection of the pre-reflection transverse wave on the plane of symmetry into zone 1 (Fig. \ref{fig:t=0}).

The regular and Mach reflections are joined at the pre-reflection contact surface where the reflected wave of the regular reflection curves across the contact surface and joins the oblique shock emanating from the Mach reflection's transverse wave. The contact surface is deflected after being shocked, curls along the wall, and jets backwards to the regular reflection.

Very prominent Kelvin-Helmholtz instabilities are present along the pre-reflection contact surface in the inviscid case (\ref{fig:r,t=17,invsicid}) and also manifest themselves in the wall jet, but they remain absent in the viscous case. The jet is also slightly longer in the inviscid case, with marginally higher pressures (not shown) at its head, but the viscous and inviscid cases are otherwise similar.
 
\section{Discussion}

While the regular reflection includes interesting features, such as the interaction between a contact surface and a shock, and its own wall jet, the discussion will focus on the Mach reflection where there are features believed to be of interest to detonation propagation.

\begin{figure}[]
\centering
\subfigure[\label{fig:p,triplepoint,t=17,viscous} Pressure]{
        \includegraphics[width=0.45\textwidth]{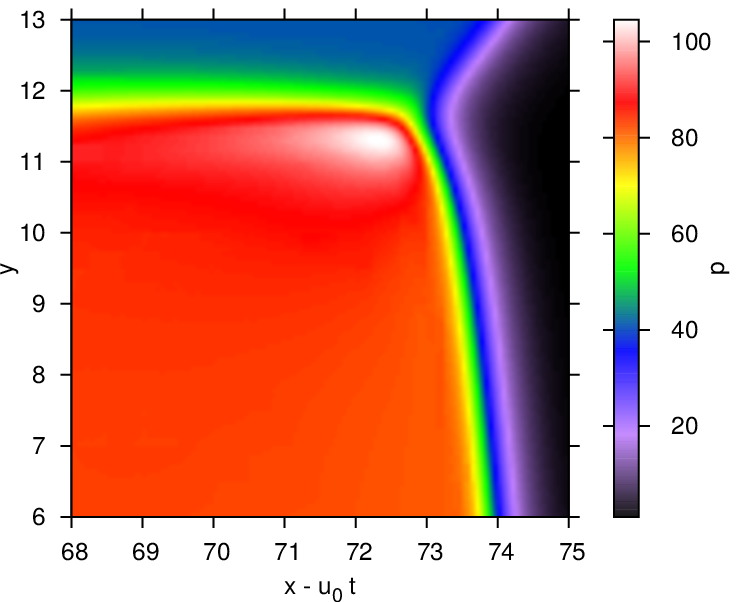}
}
\subfigure[\label{fig:r,triplepoint,t=17,viscous} Density]{
        \includegraphics[width=0.45\textwidth]{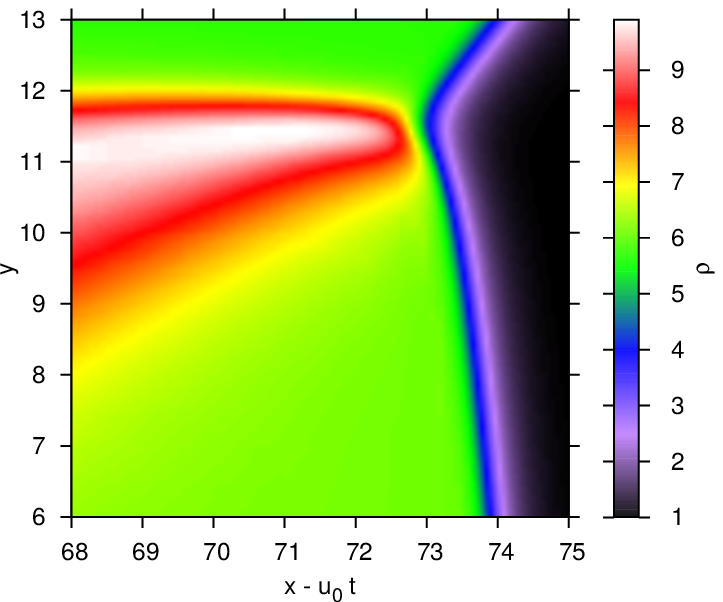}
}
	\caption{\label{fig:results_triplepoint} Curvature of shocks in vicinity of triple point, viscous, $\mathrm{Re} \approx 110$, $t=17$}
\end{figure}

One notable difference between the inviscid and viscous case is the curvature of the Mach stem, contact surface, and transverse shock in the vicinity of the triple point shown in Fig. \ref{fig:results_triplepoint} with a length scale greater than the shock thickness. A high pressure point is present behind the transverse shock in Fig. \ref{fig:p,triplepoint,t=17,viscous} but differences in density (Fig. \ref{fig:r,triplepoint,t=17,viscous}) and temperature are more subtle. Away from the triple point, the orientation of the reflected shock and contact surface differ from the inviscid case by a few degrees. The curvature and pressure concentration near the triple point persist later in time.

The angle of the viscous triple point trajectory lies above the three shock theory prediction of $5.36^{\circ}$ and eventually increases to become larger than that of the inviscid simulation later in time. The trajectory measured from the experiment in the third frame of Fig. \ref{fig:experiments} is $7.85^{\circ}$, however the gas behind the Mach stem in the experiment is likely already reacted so it does not offer a good comparison.

Following the contact surface away from the triple point, the Kelvin-Helmholtz instabilities, postulated to increase reaction rates in detonations \cite{bhattacharjee_experimental_2013,maley_influence_2015}, are absent at the induction time ($t \approx 17$). 
Traces of Kelvin-Helmholtz instabilities remain absent later into the simulation ($t \approx 120$), but appear in the inviscid simulation with regions of slightly increased temperature.

Rikanati \textit{et al.} \cite{rikanati_shock-wave_2006,rikanati_secondary_2009} measured shear layer growth rates in shock reflection experiments and found that these growth rates agreed with previous measurements for turbulent shear layers only above a Reynolds number of $\mathrm{Re} \gtrsim 2 \times 10^4$. They associated this critical Reynolds number with the transition to turbulence in Kelvin-Helmholtz instabilities, where the instabilities begin to play an important role in contact surface thickness. They defined the Reynolds number as
\begin{equation}
\mathrm{Re} = \frac{U_{\mathrm{shear}}  h_{\mathrm{M}}}{\nu_{\mathrm{c.s.}}} \notag
\end{equation}
where $U_{\mathrm{shear}}$ is the velocity difference across the contact surface, $\nu_{\mathrm{c.s.}}$ is the average dynamic viscosity across the contact surface, and $h_{\mathrm{M}}$ is the height of the Mach stem. The ideal triple-point reflection can be used to approximate the values of Reynolds number, with the Mach stem height calculated as 
\begin{equation}
h_{\mathrm{M}} = \tan(\chi) M_{\mathrm{M}} c_0 t_{\mathrm{post}} \notag
\end{equation}
where $\chi$ is the angle between the triple-point trajectory and the horizontal (in the lab frame of reference), $M_{\mathrm{M}}$ is the post-reflection Mach stem strength, and $t_{\mathrm{post}}$ is the time since reflection. This gives Reynolds numbers of $\mathrm{Re} \approx 100$ at $t=17$, when ignition is expected behind the Mach stem, and $\mathrm{Re} \approx 800$ at $t=120$ for ignition behind the transverse shock. This is two orders of magnitude smaller than the critical Reynolds number necessary for Kelvin-Helmholtz instabilities to contribute to contact surface thickness.

This estimation is extended to stoichiometric detonations under atmospheric conditions in table \ref{tab:atmospheric_detonations}, which lists the Reynolds numbers when ignition occurs behind the Mach stem, when Kelvin-Helmholtz instabilities become important \cite{rikanati_shock-wave_2006,rikanati_secondary_2009}, and at the maximal cell width. The ignition Reynolds numbers were calculated for pre-reflection incident wave strengths of $70\%$ \cite{bhattacharjee_experimental_2013} of the ideal one-dimensional steady state (Chapman-Jouguet) velocity at the end of the detonation cell, using the Shock \& Detonation Toolbox \cite{browne_numerical_2015} for Cantera \cite{goodwin_cantera_2016} with the GRI 3.0 mechanism \cite{smith_gri_1999}. Angles between the incident and Mach shocks were implied by soot foil records of ethyne (acetylene) and hydrogen detonations at sub-atmospheric pressures \cite{strehlow_strength_1969,austin_role_2003}, which showed triple-point trajectories of $\chi \approx 35^{\circ} \pm 5^{\circ} $ before reflection. This is comparable to the methane experiment presented in Fig. \ref{fig:experiments} where $\chi \approx 43^{\circ}$. The cellular Reynolds numbers were calculated using the half-cell size as the Mach stem height, approximated from Shepherd's detonation database \cite{kaneshige_detonation_1997,moen_detonation_1984,knystautas_measurement_1984,beeson_detonability_1991,bull_detonation_1982,stamps_influence_1991}.

\begin{table}[]
\centering
\caption{\label{tab:atmospheric_detonations}Reynolds numbers at ignition behind the Mach stem, at the critical value for Kelvin-Helmholtz instabilities, and for the cell size; stoichiometric atmospheric detonations ($\hat{T}_{\mathrm{ref}}=300$ K, $\hat{p}_{\mathrm{ref}}=100$ kPa, $3.72$ $\frac{\mathrm{mol\ } \mathrm{N}_2 }{\mathrm{mol\ } \mathrm{O}_2}$) of select fuels with pre-reflection incident shock Mach numbers $M_{\mathrm{I,pre}} = 70 \% M_{\mathrm{CJ}}$ and pre-reflection triple-point trajectories $\chi$}
\begin{tabular}{c|c|c|c|c|c}
 & \multicolumn{2}{c|}{$\mathrm{Re}_{\mathrm{ign}}$}  & \multirow{1}{*}{$\mathrm{Re}_{\mathrm{KH}}$} & \multicolumn{2}{c}{$\mathrm{Re}_{\mathrm{cell}}$} \\
 $\chi$ &  $40^{\circ}$ & $30^{\circ}$ & \multirow{7}{*}{$2 \times 10^{4}$} & $40^{\circ}$ & $30^{\circ}$ \\
\hline \rule{0pt}{2.2ex} CH$_4$ & 170 & 3700 &  & $1 \times 10^{7}$ & $6 \times 10^{6}$
\\ \rule{0pt}{2.2ex} 
C$_2$H$_2$ & 170 & 1300 &  & $5 \times 10^{5}$ & $3 \times 10^{5}$
\\ \rule{0pt}{2.2ex} 
C$_3$H$_8$ & 320 & 1200 &  & $4 \times 10^{6}$ & $2 \times 10^{6}$\\  \rule{0pt}{2.2ex} 
H$_2$ & 350 & 3400 &  &  $2 \times 10^{5}$  & $9 \times 10^{4}$ \\ 
\hline 
\end{tabular} 
\end{table}

The table suggests that Kelvin-Helmholtz instabilities will not appear before auto-ignition of the gas behind the Mach stem since $\mathrm{Re}_{\mathrm{KH}} \gg \mathrm{Re}_{\mathrm{ign}}$. Instabilities may arise before ignition of gas shocked by the transverse wave in methane or hydrogen reflections, as ignition there is an order of magnitude slower, and they will most likely appear within the life time of the cell because $\mathrm{Re}_{\mathrm{cell}} \gg \mathrm{Re}_{\mathrm{KH}}$.

Massa \textit{et al.} \cite{massa_triple-point_2007} performed simulations and a linear stability analysis of the contact surface and found that Kelvin-Helmholtz instabilities were attenuated by reactions (at low activation energy, not necessarily the case for stoichiometric methane-oxygen) and found diffusion to be important. This suggests instability growth will be delayed if ignition occurs before they become well established, but only momentarily until larger scales are reached as evidenced by the instabilities seen along the contact surfaces in frame 1 of Fig. \ref{fig:experiments}.

It is therefore unlikely that Kelvin-Helmholtz instabilities play a role in the propagation of detonation waves at times commensurate with the ignition time of gas behind the Mach stem, however they are likely to appear at scales on the order of the cell size where they may accelerate reactions. These instabilities may also be important in problems such as detonation re-initiation where the shocks are much weaker and induction time is longer.

The contact surface curves towards the Mach stem as it approaches the wall, caused by flow stagnation perpendicular to the reflecting surface \cite{li_reconsideration_1995,li_analysis_1999,mach_bifurcating_2011}. The stagnation pressure to the left of the jet (not shown) increases with the Reynolds number as viscous losses along the contact surface become less significant. This is accompanied by the development of a vortex, which may help with mixing, but the wall jet has grown little and not caused the Mach stem to bulge. The jet and vortex remain much weaker than their inviscid counterparts. Temperature in the wall jet drops along the contact surface, possibly reducing the effect of combustion radicals, while the temperature behind the Mach stem increases, decreasing ignition time late in the reflection process. Heating of the contact surface \cite{lau-chapdelaine_viscous_2015} does not occur due to the inclusion of heat diffusion in the model. 

The use of a better model for viscosity, including the bulk viscosity for example, would lower the simulation's Reynolds number, despite the difference of Mach stem height with three shock theory, and amplify differences between the viscous and inviscid cases.

The Mach reflection begins to adopt the features of the inviscid case (of earlier times) as the Reynolds number increases, which highlights the importance of including viscosity in the simulation of time-limited processes such as detonation.

\section{Conclusion}
The reflection of the triple-shock configuration was studied numerically in two dimensions using the Navier-Stokes equations under conditions similar to those present in the creation of detonation cells. The reflection of a triple point on a plane of symmetry creates a double Mach reflection, from the reflection of the incident Mach stem, and a backwards-facing regular reflection occurs from the reflection of the transverse shock.

When viscosity was considered, no Kelvin-Helmholtz instabilities were seen and there was no bifurcation of the Mach stem for Reynolds numbers corresponding to the ignition time behind the Mach stem and transverse shock. However, a small wall jet was present, which increased with Reynolds number and eventually formed a small vortex. Kelvin-Helmholtz instabilities are unlikely to be a source of rapid reactions at early times in detonation propagation. Further study is required on longer time scales, including those associated with the cell size.

This study shows that viscosity has an important role on some shock reflection mechanisms believed to accelerate reaction rates in detonations. Diffusion must be considered when the time scales are small, and the Euler equations are inadequate in this regard.

\begin{acknowledgements}
This work was sponsored by a Natural Sciences and Engineering Research Council of Canada (NSERC) Discovery grant to M. I. Radulescu and a NSERC Alexander Graham Bell Canadian Graduate Scholarship to S. S.M.
Lau-Chapdelaine.
\end{acknowledgements}

\bibliographystyle{splncs} 
\bibliography{template.bib}   

%
%

\end{document}